\documentclass{aa}
\usepackage[varg]{txfonts}
\usepackage{graphicx}

\usepackage[normalem]{ulem}

\begin{document}

\title{WD 1145+017: Alternative models of the atmosphere, dust clouds, and gas rings}
\titlerunning{WD 1145+017: alternative models}
\author{J.~Budaj\inst{1} \and A.~Maliuk \inst{1} \and I.~Hubeny \inst{2}}
\institute{Astronomical Institute, Slovak Academy of Sciences,
05960 Tatransk\'{a} Lomnica, Slovak Republic\\
\email{budaj@ta3.sk, amaliuk@ta3.sk}
\and The University of Arizona, Steward Observatory, 933 North Cherry Avenue, Tucson, AZ 85719, USA}
\date{Received ???? ??, ????; accepted ???? ??, ????}

\abstract
{WD 1145+017 (WD1145) is the first white dwarf known to be orbited by disintegrating exoasteroids.  It is a DBZ-type white dwarf with strongly variable broad circumstellar lines and variable shallow  ultraviolet (UV) transits. Various models of the dust clouds and gaseous rings have been proposed as an explanation for this behavior.}
{We aim to revisit these observations and propose alternative or modified
models of the atmosphere of this white dwarf, its dust clouds, and gas rings.}
{The simple radiative transfer code Shellspec was modified for this purpose and used for
testing the new dust cloud and gas disk models. We used modified TLUSTY and SYNSPEC codes to calculate atmosphere models assuming the local thermodynamical equilibrium (LTE) or nonLTE (NLTE), and to calculate the intrinsic spectrum of the star.
We then used these atmosphere models to estimate the mass of the radiative and convective zones and NLTE spectrum synthesis to estimate their chemical composition.}
{We offer an alternative explanation of some (not all) shallow UV transits. These may be naturally caused by the optical properties of the dust grains: opacities and mainly phase functions as a result of the forward scattering. The latter is much stronger in UV compared to the optical region, leaving more UV photons
in the original direction during the transit. We also developed an alternative model of the gaseous disk, consisting of an inner, hotter, and almost circular disk and an outer, cooler, and eccentric disk. The structure precesses with a period of $3.83 \pm 0.12$ yr. We demonstrate that it fits the observed circumstellar lines reasonably well. These alternative models solve a few drawbacks that might be associated with the previous models, but they also have their own disadvantages.  We confirm that the chemical composition
of the atmosphere is similar to that of CI chondrites but carbon, nitrogen, and sulfur are significantly underabundant
and much closer to the bulk Earth composition. This is a strong argument that the star has recently encountered and accreted material from a body of Earth-like composition.}
{}

\maketitle

\section{Introduction}

It is well known that some white dwarfs may have infrared (IR) excess \citep{zuckerman87}. 
Also, a large fraction of white dwarf atmospheres (between one-quarter and one-half) are 
polluted with heavy elements \citep{zuckerman03,zuckerman10,koester14}.
These heavy elements should have quickly settled down and been depleted from the atmosphere because of the strong gravity at the surface of these stars \citep{paquette86,koester09}.
It was suggested \citep{debes02,jura03} that this IR excess and metallicity of white dwarfs
might be explained by perturbations of the original planetary orbits that may bring planets
or smaller bodies closer to the white dwarf where they are tidally disrupted into dust debris which can accrete and contaminate the surface of the star.
Subsequent observations \citep{kilic06,hippel07,jura07} confirmed this suggestion. 
The abundance ratios of heavy elements in the atmospheres of some 
white dwarfs were found to closely match the abundance patterns in rocky 
bodies in the Solar System \citep{zuckerman07}.
A more detailed road map towards polluted white dwarfs with an IR excess was proposed recently by \cite{brouwers22}, which begins with the tidal disruption of an asteroid, followed by the formation of a highly eccentric tidal disk, and then collisional grind-down of fragments and their subsequent circularization and accretion.

The most direct proof of this scenario was the discovery of a white dwarf WD1145+017 (WD1145),
also known as EPIC 201563164 \citep{vanderburg2015}. The star exhibits IR excess and strong lines of Mg, Al, Si, Ca, Fe, and Ni.
Photometric observations revealed variable asymmetric eclipses of up to 40\% depth
 with periods from 4.5 to 4.9 hours. These periods correspond to a distance of about 1 $R_{\odot}$ from the star.
The interpretation made by the authors is that the star is orbited by one or more disintegrating
asteroids on close orbits, producing dusty comet-like tails that are responsible for the eclipses.
A follow-up study by \cite{gansicke16} revealed that the WD1145 system has dramatically evolved since 
its discovery.  Multiple transit events have been detected
in every light curve with variable duration of 3$-$12 minutes and depths 
of 10$-$60\%. The shortest-duration transits require the occulting cloud of debris to be several times the size of the white dwarf.
This confirmed the conclusion of \cite{vanderburg2015} that the transits 
are not caused by solid bodies, but by clouds of gas and dust flowing from 
tiny objects that remain undetected. 
\cite{rappaport16} and \cite{rappaport18a} carried out
long-term monitoring of the photometric variability of the system during 2015-2017. The optical activity of the system and eclipses were found to be increasing 
in strength and duration. The dominant period was about 4.49 hours but there were numerous (15) smaller dips that drifted systematically in phase with respect to the main dip period.

\cite{alonso2016} and \cite{izquierdo18} used the 10m GTC telescope and carried out observations in the wavelength range from 480 to 920 nm and split them into four or five bands.
These authors found that the eclipses in all bands are 
consistently the same, which enabled them to set a lower limit on the dust particle 
sizes of about 0.5 $\mu$m.
Subsequent simultaneous IR and optical observations of  WD 1145 by \cite{zhou2016} over the wavelength range of 0.5-1.2 $\mu$m confirmed
the result. These latter authors revealed no measurable difference in transit depth for multiple photometric passbands.
This result allowed the authors to rule out (with 2$\sigma$ significance) particles 
smaller than 0.8 $\mu$m and favor the minimum particle size 
of about $a_{\rm min}$ = 10$^{+5}_{-3}$ \, $\mu$m.
\cite{xu2018} reported two epochs of multi-wavelength photometric 
observations of WD 1145, including several filters in the optical, 
K$_s$, and 4.5 $\mu$m bands. 
These authors also found that the transit depths are the same at all wavelengths, at least to within the observational uncertainties. The authors concluded that their non-detection of
wavelength-dependent transit depths from optical to 4.5 $\mu$m
implies that the transiting material around WD 1145
must mostly consist of grains of sizes $\ge$ 1.5 $\mu$m.
Multi-wavelength ground-based photometry in V and R bands by \cite{croll17} did not reveal any difference in transits, which means that the grains
are either larger than 0.15 microns or smaller than 0.06 microns.
Surprisingly, first \cite{hallakoun17} in the optical region and then \cite{xu2019}
in ultraviolet (UV) found that the UV transit depths are always shallower than those in the optical. The results of these authors and their explanation are described in more detail in Sect. \ref{xumodel}.

\cite{xu16} discovered numerous broad absorption lines in the spectra of WD1145 which are due to circumstellar material.
\cite{redfield17} show that these lines are variable on timescales of minutes to months and suggested an eccentric disk model as an explanation.
The strength of these lines decreases during eclipses.
\cite{cauley18} revealed that over the course of 2.2 years these lines show complete velocity reversal from redshifted to blueshifted 
and proposed a model of 14 eccentric gas rings undergoing relativistic precession.
This model was further supported and elaborated by \cite{fortin20} and is described in more detail in Sect. \ref{oldringmodel}.

Since then, further indications have been found that asteroids, planets, or their debris  also orbit other white dwarfs : ZTF J013906.17+524536.89 \citep{vanderbosch20},
SDSS J122859.93+104032.9 \citep{manser19}, WD J091405.30+191412.25 \citep{gansicke19}.
Most recently, a Jupiter-sized planet was directly detected transiting WD 1856+534 \citep{vanderburg20}.
 
In the present paper we aim to revisit the existing photometric and spectroscopic observations of WD1145 and propose alternative models for their explanation. In Sect. \ref{star} we deal with the stellar atmosphere
model and its chemical composition. Section \ref{transit} is devoted to transits and dust clouds. Section \ref{variab} describes the variability
of circumstellar lines and Sect. \ref{gasdisk} presents our gas disk model.

\section{Observational data}

We made use of photometric and spectroscopic data from various sources.
Spectroscopic data obtained with the HIRES instrument on the KECK telescope \citep{vogt94}
were taken from the Keck Observatory Archive. These have a resolution of about 40\,000. 
Spectroscopic data from the X-SHOOTER instrument \citep{vernet11} on the VLT telescope
were taken from the ESO Science Archive Facility, and have a resolution of about 7000. 
Observations in the UV region were made with the NASA/ESA Hubble Space Telescope and were obtained from the data archive at the Space Telescope Science Institute.
These have a resolution of about 20\,000.
These data are described in more detail in \cite{xu16,cauley18,xu2019}.

Measurements of the transit depth in different filters were taken from \cite{xu2019}.
The GAIA DR3 data release \citep{gaia16,gaiadr3_21} lists a parallax of 
$6.874\pm 0.091$ mas, which translates to a distance of $145.5\pm 1.9$ pc. A small
offset of 0.021 mas \citep{groenewegen21} was already incorporated into this distance estimate.

\section{Star}
\label{star}

\subsection{Radiative transfer in the circumstellar dust and gas}

Our models consist of the white dwarf and circumstellar material (CM)
orbiting the star in the form of dust and gas. 
We calculated the light curves and spectra of this system (model)
using the code Shellspec \citep{budaj04}.
The code solves a simple radiative transfer along the line of sight
in the circumstellar medium assuming local thermodynamic equilibrium (LTE) and single scattering approximations.
As a boundary condition for the radiative transfer in the CM, we used spectra 
obtained from the stellar atmosphere models. These are described in the separate 
Sect. \ref{atm}.

\subsection{Stellar atmosphere model}
\label{atm}

An atmosphere model is characterized by its effective temperature and surface gravity.
These might be derived from the spectral energy distribution (SED) or spectral lines originating from different excitation potentials or ionization stages.
However, UV fluxes are sensitive to nonLTE (NLTE) effects, dust extinction, and, in this case, to absorption due to gaseous rings. Most of the spectral lines are blended and are sensitive to elemental abundances or saturation effects, for example due to microturbulence. Helium lines are an exception because they are broad and He is by far the most abundant element in the atmosphere. This means that its lines are not sensitive to the abundance but are sensitive mostly to the temperature. Hence, we adopted the effective temperature of $T_{\rm eff}=15\,000$ K which  best fits the strong He lines in the optical region. This value is in agreement with \cite{xu2019}. We also adopted a canonical value for the surface gravity $\log g=8.0$ in cgs units, which is also in agreement with the above-mentioned paper. 
A turbulent velocity of 3 km\,s$^{-1}$ and zero rotational velocity were assumed. 
To calculate the intrinsic spectrum of the white dwarf,
we first calculated the  LTE model of the stellar atmosphere using the code TLUSTY209 \citep{hubeny88,hubeny95,hubeny21}.  
TLUSTY models are plane-parallel, horizontally homogeneous (1D)
models in a hydrostatic and radiative--convective equilibrium and include a line blanketing effect. The LTE
model determines the basic atmospheric structure (temperature, mass
density, and electron density as functions of depth). We further assume that these state quantities are fixed and calculate  NLTE atomic level populations 
self-consistently with the radiative transfer equation using the same code.
This is a so-called restricted NLTE model of the stellar atmosphere.
For this purpose, it is necessary to provide models of the atoms, which we took from \cite{lanz07}. We considered dozens of atomic levels of \ion{H}{I-II}, \ion{He}{I-III}, \ion{C}{I-III}, \ion{O}{I-III}, \ion{Mg}{II-III}, \ion{Al}{II-IV}, \ion{Si}{II-IV}, and \ion{Fe}{I-IV} in NLTE.

Unlike previous model atmospheres computed by TLUSTY, we consider here
level dissolution  for all helium energy levels considered explicitly,
together with additional opacity due to pseudo-continua, following the
formalism developed by \cite{hubeny94}. This theory was developed
for hydrogen and hydrogenic ions, but because the effects of dissolution
are important particularly for high-energy levels, we believe that
using the hydrogenic approximation for neutral helium represents a
reasonable approximation.
Similarly, we implemented the hydrogenic level dissolution for helium
in Synspec.

\begin{figure}[ht]
\centerline{\includegraphics[width=0.45\textwidth]{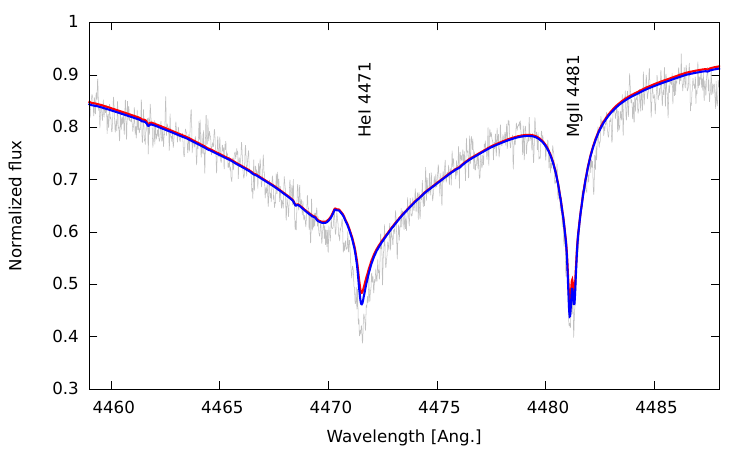}}
\centerline{\includegraphics[width=0.45\textwidth]{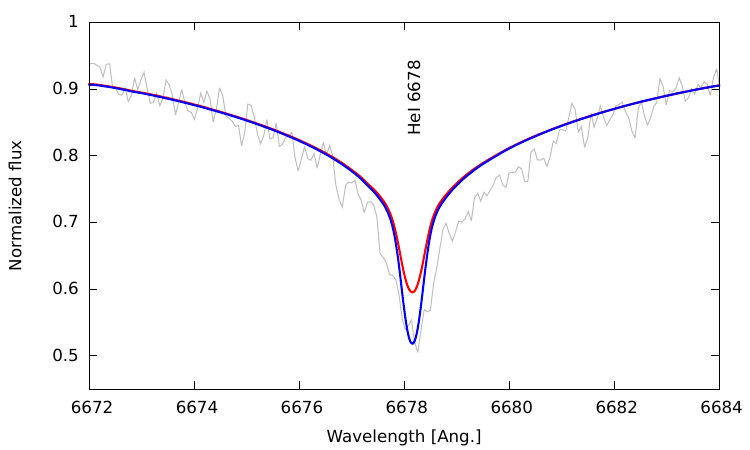}}
\caption{Theoretical and observed spectra in the vicinity of \ion{He}{I} 4471 and \ion{Mg}{II} 4481 lines (top) and \ion{He}{I} 6678 line (bottom). Helium and magnesium lines in NLTE (blue) have slightly deeper cores than in LTE (red). Observations are plotted in gray.}
\label{he4471}
\end{figure}

We find these NLTE effects to be important for the spectrum mainly
in the UV. The Lyman continuum is almost 30\% stronger in NLTE and the continuum below the \ion{He}{I} ionization edge is about two orders of magnitude stronger in NLTE.
NLTE also heavily affects strong \ion{He}{I} lines in UV and makes cores of \ion{He}{I} lines in the optical region slightly deeper. This is shown in Fig. \ref{he4471}. The chemical composition of the star was initially taken from \cite{xu16}. However, in our synthetic spectra, most of the metal lines are stronger than the observed ones, and so we adjusted the abundances to fit the stellar lines. These abundances are discussed separately below in Sect. \ref{abundances}.

\begin{figure*}[ht]
\centerline{
\includegraphics[width=0.45\textwidth]{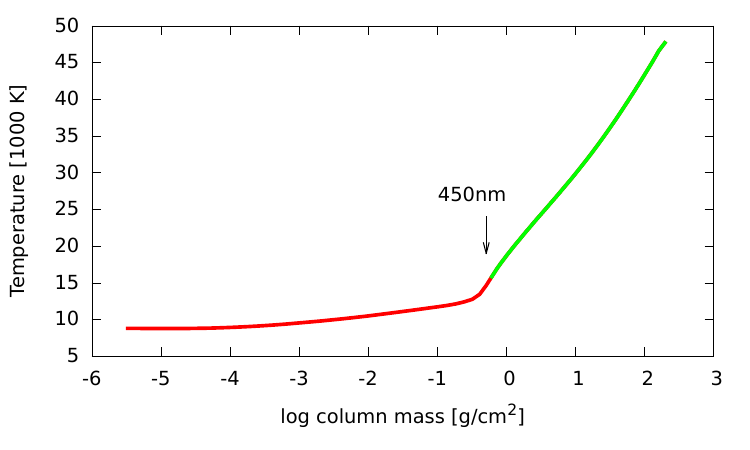}
\includegraphics[width=0.45\textwidth]{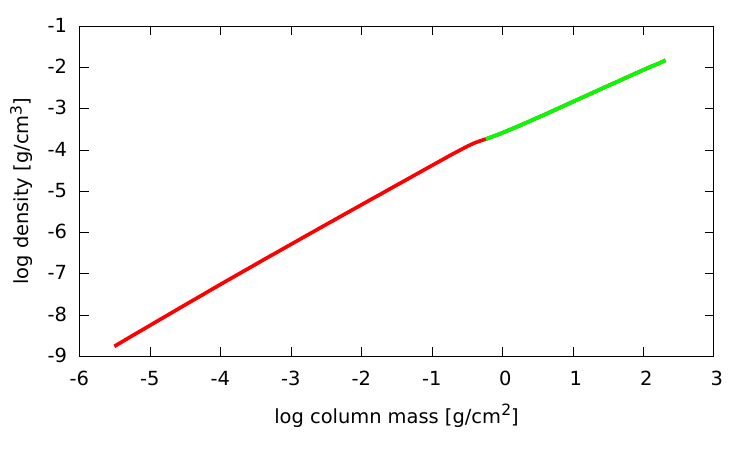}}
\caption{Temperature (left) and density (right) in the atmosphere
as a function of the column mass. Convective regions are marked
in green. An arrow indicates the optical depth of about unity in the continuum at 450 nm.}
\label{temp_m}
\end{figure*}

Deeper layers of the atmosphere are convective. To pursue this convection deeper, we calculated deeper models in pure LTE with the same code, including a few highly ionized explicit elements,
namely \ion{H}{I-II}, \ion{He}{I-III}, \ion{O}{I-V}, and \ion{Fe}{II-V}. These models made use of a pre-calculated opacity table
which includes the line blanketing effect from all elements and assumes a chemical composition which is  described in the following section. Models span the column masses from less than $10^{-5}$
up to more than $10^{2}$ g\,cm$^{-2}$ and densities from $10^{-9}$ up to more than $10^{-2}$ g\,cm$^{-3}$.
The behavior of the temperature and density as functions of the column mass
are shown in Fig. \ref{temp_m}.
The column mass is the mass of a column of the atmosphere above a unit surface located at a certain depth in g\,cm$^{-2}$. This quantity is often used as a depth variable.
One can observe that a fraction of the atmosphere
at column masses greater than about 1 g\,cm$^{-2}$ is convective. This is illustrated in the figures by a different color
when the convection carries more than 10\% of the flux.
The onset of convection in the models is quite sharp and our results
do not change much if we adopt a different threshold.
By convection, we understand convective regions where 
the classical (Schwarzschildt) condition for convection is satisfied \citep{hubeny14} and convection transports a significant amount of energy.
This does not mean that the non-convective (radiative) layers above are void of any motion. One might expect overshooting, sound waves, turbulence, and other phenomena that 
might still mix the medium but do not transport a lot of energy.
The upper layers are optically thin and the free escape of photons
and heat is much more efficient in transporting the energy than the convection.
This may have implications for the observed chemical composition
of the star. Accretion of metals would mainly contaminate these upper, stable, nonconvective layers where they are observed. As soon as the metals sink into the convection zone, they become mixed efficiently with the material in the convection zone. However, as mentioned above, these upper, nonconvective layers may not be stable and may be mixed with the convection zone too. In this case, the accretion would have to also contaminate the whole convection zone. Estimates of the mass of the convection zone differ. \cite{xu16} estimated that it currently contains $6.6\,10^{23}$ g of heavy elements while \cite{fortin20} list
about $1\,10^{23}$ g of metals.
Our deepest models can put a lower limit on the mass of the convection zone of $2\,10^{21}$ g
or  $3\,10^{-7} M_{\oplus}$.
Our models may also be used to estimate the mass of the upper nonconvective region of the atmosphere.
 Assuming the radius of the star to be $R_{\star}=0.012 R_{\odot}$, the mass of the nonconvective fraction of the atmosphere is about $5\,10^{18}$ g or $9\,10^{-10} M_{\oplus}$. This is much less than the mass of the convection zone. 
 
 A lower constraint on the mass of the contaminated atmosphere might be set
by the spectral line formation. The chemical composition is derived from 
 the metal lines in the optical region. This means that the observed chemical
 composition reflects the composition at those layers where
these spectral lines are formed. That is why we also show in 
Fig. \ref{temp_m} the depth corresponding to the optical
depth of about two-thirds in the continuum at 450 nm. This corresponds to the B filter. As we can see, this depth is very close to the top of the convective zone. This is not simple incidental;
as mentioned above, the convection stops operating as soon as the medium becomes optically thin.
This is why the depth of spectral line formation gives a lower constraint on the mass of contaminated atmosphere that is equal to the mass of upper nonconvective layers,
that is, it must be greater than $5\,10^{18}$ g.

After calculating the model of the atmosphere, we calculated the spectrum
emanating from this model using the code SYNSPEC54 \citep{hubeny17}.
This spectrum was then used as input (boundary condition for the radiative transfer in CM) for the code Shellspec.

\subsection{Chemical composition of the atmosphere}
\label{abundances}

As a byproduct of our analysis, we are also able to briefly explore the chemical composition of the atmosphere of the white dwarf, which was analyzed before by \cite{xu16} and \cite{fortin20} assuming LTE. 
Our study differs from their analyses by including NLTE effects, which nevertheless turned out to be of relatively minor importance for the abundances.

The observed spectrum features absorption lines with a narrow component and a broad one. The narrow component originates from the star
while the broader component originates from the circumstellar material.
The stellar component is shifted by about 43 km\,s$^{-1}$ and this shift is, to a large extent, due to the gravity reddening of the star as already pointed out by \cite{xu16}.
Our stellar abundances are mainly based on the spectrum from the optical region. However, UV lines enabled us to estimate the abundance of aluminum (\ion{Al}{II} 1539.8, 1670.8), sulfur (\ion{S}{II} 1204.3, 1253.8, \ion{S}{I} 1295.7, 1425.0), and nickel (numerous lines). A few lines of a new element, phosphorus (\ion{P}{II} 1149.9, 1155.0, 1157.0, 1159.1, 1249.8), were identified and its abundance determined. We roughly estimated the abundance of carbon based on
\ion{C}{II} 1334.5, 1335.7 lines, put constraints on the oxygen abundance from \ion{O}{I} 1152.1, 1302.2, 1304.9, 1306.0 (the last three lines are strongly affected by geocoronal line emission), and put an upper limit on the nitrogen abundance based on \ion{N}{I} 1135.0, 1199.6, 1200.7 lines. Silicon abundance based on some strong UV lines seems to be a factor of two lower than that derived from the optical region and we assumed the latter to be more reliable.

\begin{figure*}[ht]
\centerline{
\includegraphics[width=0.45\textwidth]{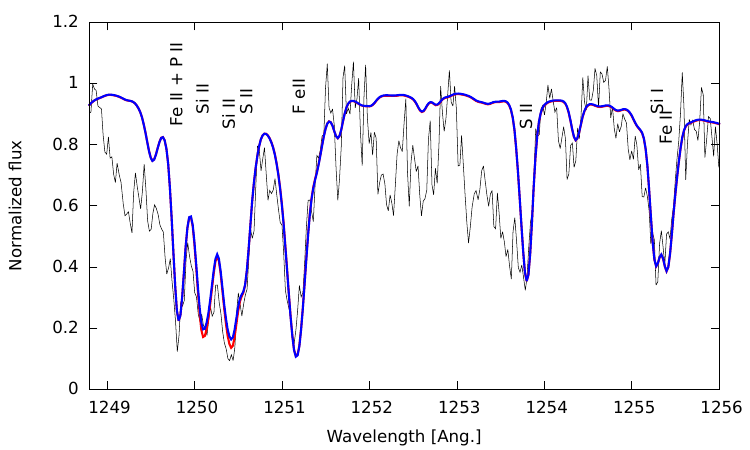}
\includegraphics[width=0.45\textwidth]{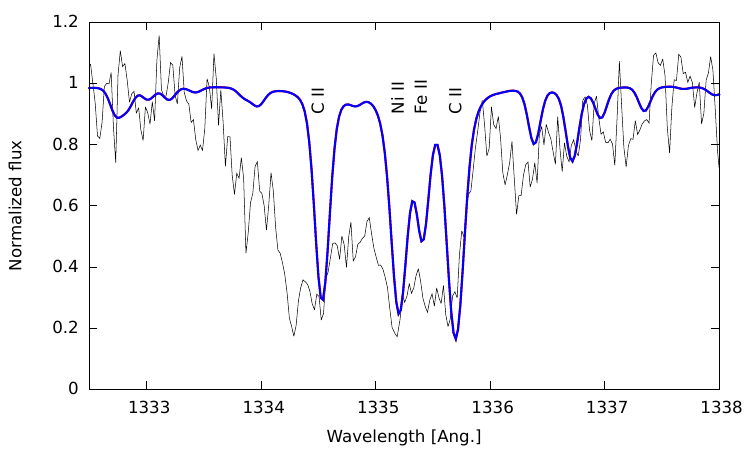}}
\centerline{
\includegraphics[width=0.45\textwidth]{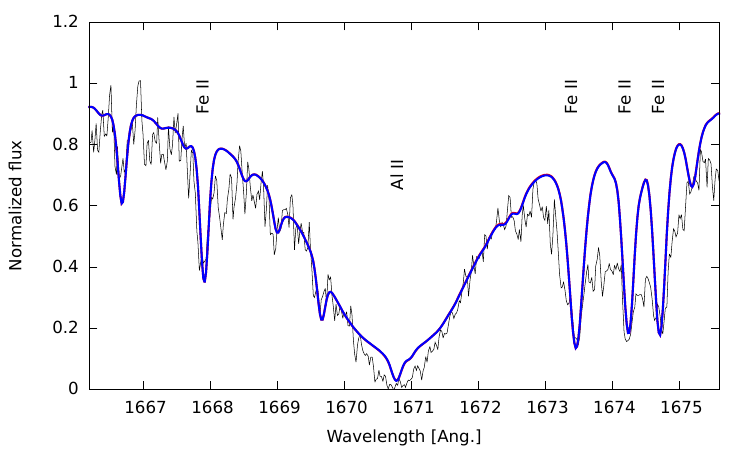}
\includegraphics[width=0.45\textwidth]{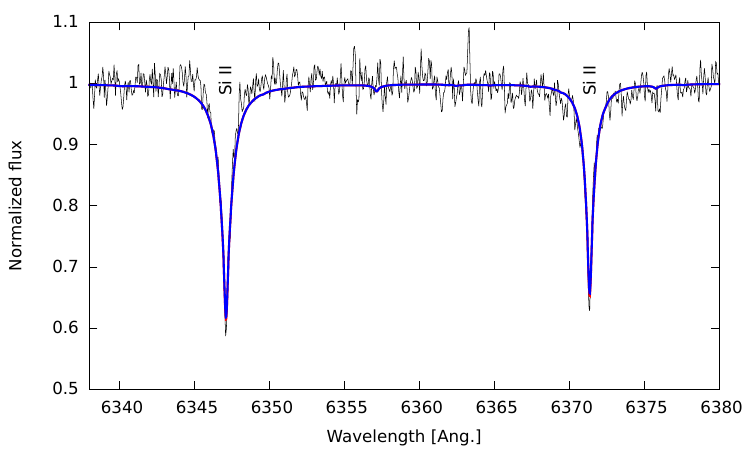}}
\caption{Comparison of synthetic spectra of the atmosphere assuming LTE (red) or NLTE (blue) with the observations (black).
Top-left: Region with the lines of sulfur. Top-right: Region containing
the lines of carbon. Bottom-left: Region with aluminum lines. Bottom-right:
Silicon lines in the optical region.}
\label{spectra}
\end{figure*}

The mass fraction of hydrogen, helium, and metals are $X=2.5\,10^{-6}$, $Y=1$, and $Z=6\,10^{-5}$, respectively. This means that the convection zone contains at least $2\,10^{-11} M_{\oplus}$ of heavy elements. Most of the elements have slightly lower abundances (by about 0.2 dex) than in \cite{fortin20}. 
See Table \ref{abund} for a comparison. This is most probably caused by differences in the atmospheric structure which is very sensitive to the treatment of line blanketing or \ion{He}{I} resonance lines.
Figure \ref{spectra} shows a comparison of observations and synthetic spectra of the star assuming both LTE and NLTE.
A few regions containing spectral lines of key elements are selected. We note that NLTE effects 
make the cores of \ion{He}{I} and \ion{Mg}{II} lines slightly deeper. Our rough estimate of the uncertainties in abundances is about 0.3 dex.

\begin{figure}[ht]
\centerline{
\includegraphics[width=0.45\textwidth]{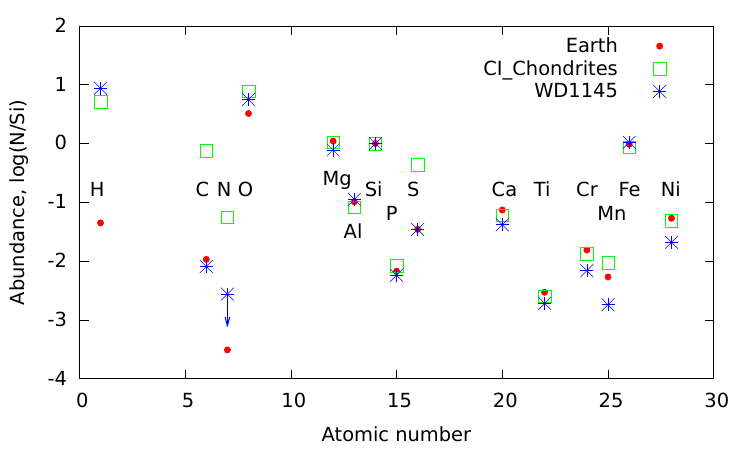}}
\caption{Photospheric abundances of WD1145 (blue stars) compared with the abundances of CI chondrites (green squares) and those of bulk Earth (red circles). Abundances of C, N, and S are close to Earth values.
Abundances were scaled to silicon abundance.}
\label{abundfig}
\end{figure}

If the enhanced metal abundances were due to accretion of gas delivered by an asteroid, the abundance pattern would be similar provided that the diffusion or gravitational settling of different elements is not highly selective \citep{paquette86,koester09}.
A recent study of abundances in the atmospheres of 202 DZ white dwarfs by \cite{harrison21}
shows that the majority (>60\%) of systems are consistent with 
the accretion of primitive material. 
Therefore, in Fig. \ref{abundfig} we show our photospheric abundances of WD1145,
the element abundances of CI chondrites \citep{asplund09}, and the bulk composition of the Earth.
All are scaled to zero silicon abundance. The chemical composition of CI chondrites is often used as a
reference because it represents primitive material from the early Solar System.
CI chondrites are also closest to the photospheric abundances of the Sun which contains most of the mass of the Solar System. The chemical composition of CI
chondrites is also linked to the bulk chemical composition of a rocky planet (Earth).
\cite{xu16} state that the chemical composition of WD1145 is akin to that in CI chondrites,
while \cite{fortin20} state that it is very similar to what might be 
expected for the accretion of a rocky body with bulk Earth composition, but their abundances provide only weak support for the claim, because the abundances of CI chondrites and bulk Earth are very similar.
It is obvious from Fig. \ref{abundfig} that the abundance pattern is similar to both CI chondrites and the bulk Earth. Even the abundance of hydrogen is not too far from that of CI chondrites and might be partially related to the same origin. However, the photospheric abundances of carbon, nitrogen, and sulfur are significantly lower than those of CI chondrites, and are much closer to the bulk Earth composition. Therefore, our abundances can be used to distinguish  between the two and present a strong argument for the second claim, that is that the star has recently encountered and accreted material from a body of Earth-like composition.

\begin{table}[ht]
\caption{
Abundance of a chemical element, N, in the atmosphere of WD1145 relative to helium by number 
$(\log(\rm N/He))$.
The second column contains results from this paper. In the third column are results from \cite{fortin20} for comparison.
The fourth column shows abundances of CI chondrites relative to silicon $(\log(\rm N/Si))$ taken from \cite{asplund09}. The first column is an element designation.}
\label{abund}
\centering
\begin{tabular}{lrrr}
\hline\hline
N  & WD1145  & FDX     & Chondrite\\
\hline
 H &   -5.00 &  -5.00  &  0.71  \\
He &    0.00 &   0.00  & -6.22  \\
 C &   -8.03 &  -7.50  & -0.12  \\
 N & $<-8.49$&$<-7.00$ & -1.25  \\
 O &   -5.19 &  -5.12  &  0.89  \\
Mg &   -6.05 &  -5.91  &  0.02  \\
Al &   -6.89 &  -6.89  & -1.08  \\
Si &   -5.94 &  -5.89  &  0.00  \\
 P &   -8.18 &      -  & -2.08  \\
 S &   -7.40 &$<-7.00$ & -0.36  \\
Ca &   -7.32 &  -7.00  & -1.22  \\
Ti &   -8.66 &  -8.57  & -2.60  \\
Cr &   -8.10 &  -7.92  & -1.87  \\
Mn &   -8.68 &  -8.57  & -2.03  \\
Fe &   -5.92 &  -5.61  & -0.06  \\
Ni &   -7.62 &  -7.02  & -1.31  \\
\hline
\end{tabular}
\end{table}

\section{Model of dusty transits}
\label{transit}

\subsection{Current explanation of the shallow UV transits}
\label{xumodel}

\cite{xu2019} performed a large photometric and spectroscopic campaign involving HST, Keck, VLT, and Spitzer. They observed the transits of dust clouds at various wavelengths.
In UV at 1300\AA,\, in optical at 4800 and 5500 \AA,\, and in IR at 4.5 microns.
Transit depths as well as transit depth ratios in different filters varied significantly over time.
Somewhat surprisingly,  these latter authors discovered that UV transits are always shallower
than those in the optical region. A following explanation of this
phenomenon was proposed.
There is an inner flat gas ring absorbing the light from the star.
It absorbs mainly in UV, because there are many strong lines, and shows no significant absorption in the optical region.
The dust cloud associated with the asteroid is orbiting farther out and is aligned
with the gas ring.
It eclipses the gas ring and the star but does not cause a deep eclipse in UV because that region was already eclipsed by the gas.  On the contrary,  it can cause a deeper eclipse in the optical because the star was not eclipsed by the gas at these wavelengths.
\cite{xu2019} also found a very small IR/optical transit depth ratio (on average equal to 
$0.235\pm 0.024$), but after correcting for the excess IR emission, this ratio was consistent with unity ($0.995\pm 0.119$).
A very mild ``bluing'' effect during transits was reported earlier by \cite{hallakoun17} using ULTRACAM on the 3.6m NTT telescope with u'g'r'i' filters and those authors offered a similar explanation. \cite{izquierdo18} arrived at the same conclusion that the gas ring and the dust cloud are aligned, because they observed a significant decrease in strength of the circumstellar lines during the transit.

\subsection{Alternative model and explanation of the shallow UV transits}

\subsubsection{Dust grain properties}
\label{dust}

We base our explanation of this phenomenon on intrinsic dust grain properties.
Dust orbiting the white dwarf is relatively close to the star
(about 1 $R_{\odot}$) which means that it gets quite hot (over 1000K)
and only refractory grains can survive under these conditions.
We consider two different types of dust: silicates and iron. Both are refractory species. Both are optically active
and abundant, and both can be found
in a big asteroid either at the surface or in its core.
We represent silicates by the `astronomical silicate' and adopt
its complex index of refraction from \cite{draine03}. 
The complex index of refraction of iron was taken from \cite{pollack94}.
We further assume spherical homogeneous grains with a 
grain size distribution of \cite{deirmendjian64}. This is a relatively narrow distribution
characterized by a modal particle radius. Unless stating otherwise, by a particle size $r$
we refer to the modal particle radius of this distribution of particle sizes expressed in microns.
We calculated scattering and absorption opacities and phase functions
for a region covering 0.1-1000 microns as described in \cite{budaj2015} using a modified version of the publicly available Mie code of \cite{bohren83}.

\begin{figure}[ht]
\includegraphics[width=9.cm]{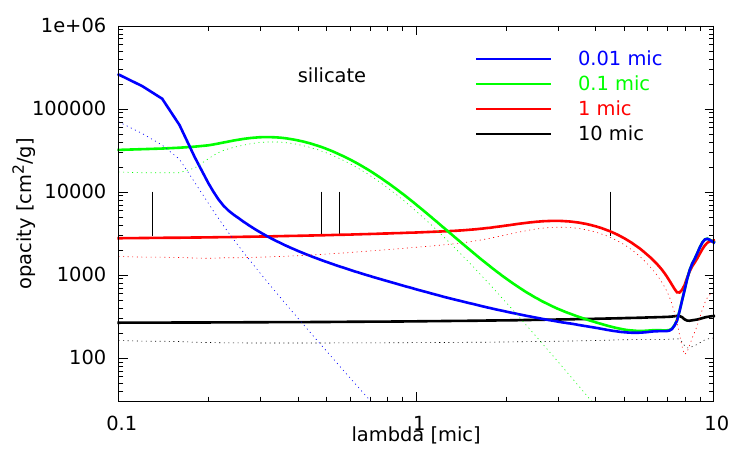}
\includegraphics[width=9.cm]{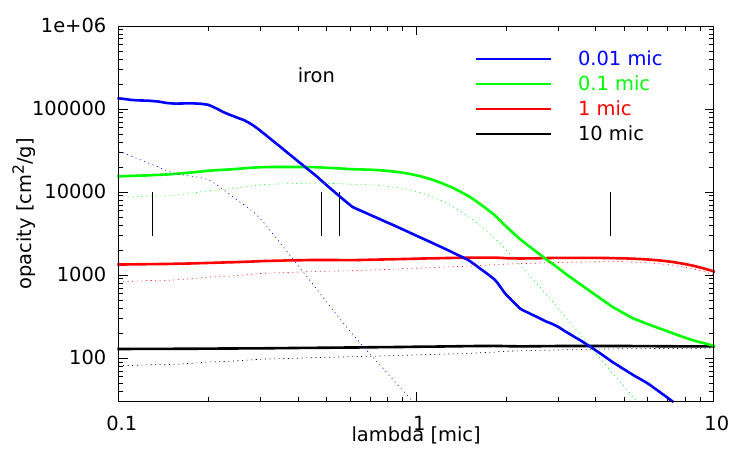}
\caption{Total (absorption and scattering) opacities (full lines)
and scattering opacities (thin dashed lines) for silicate (top)
and iron (bottom) for a few particle sizes. Vertical lines show the position of filters used for the observations at 0.13, 0.48, 0.55, and 4.5 microns.}
\label{fig:opac}
\end{figure}

\begin{figure}[ht]
\includegraphics[width=0.49\textwidth]{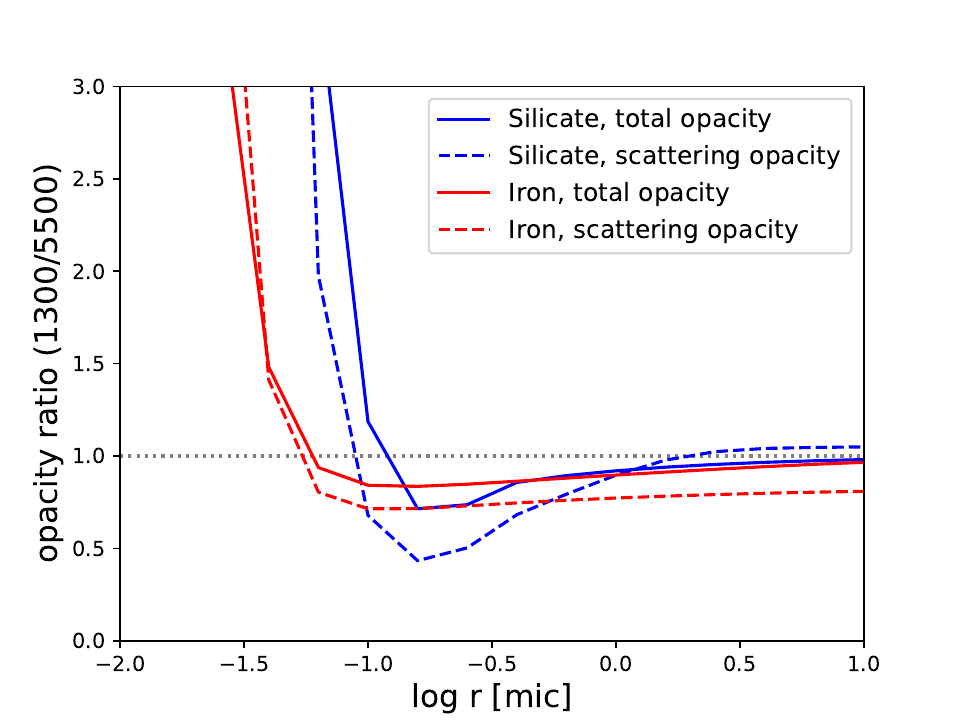}
\includegraphics[width=0.49\textwidth]{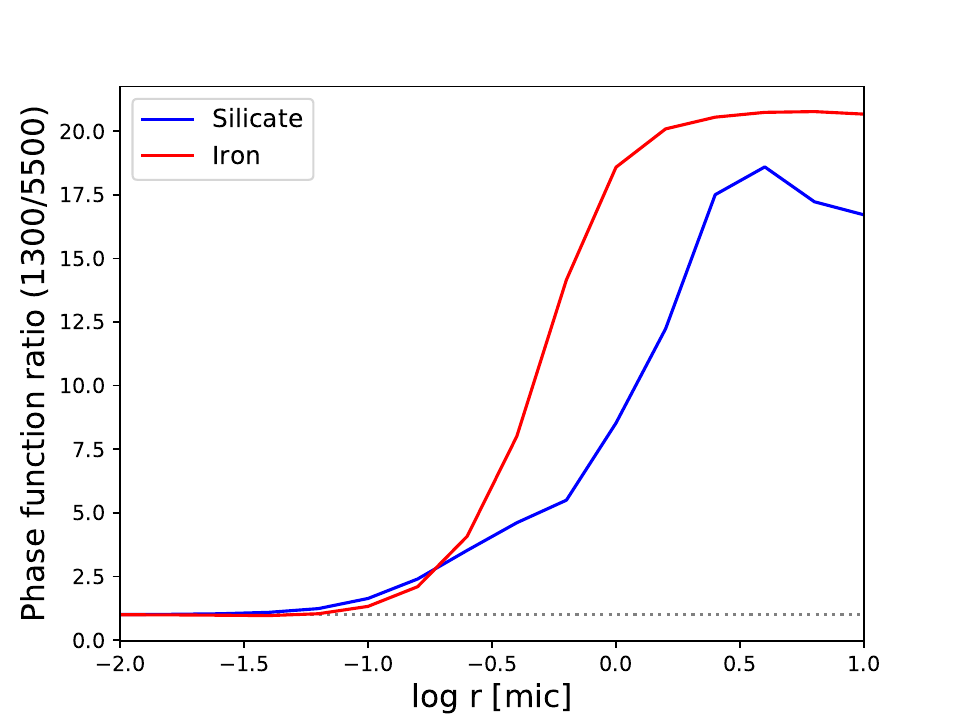}
\caption{Top: Ratio between opacities at 1300\AA\ and 5500\AA\ for silicate (blue) and iron (red). Solid lines correspond to ratios of total opacity, dashed lines correspond to ratios of scattering opacity. Values smaller than one could cause shallow UV transits.
Bottom: Ratio of the phase function at 1300\AA\ and 5500\AA\ for astronomical silicate (blue) and iron (red). Calculated for the scattering angle zero, i.e., for
the forward scattering. Values larger than one might also cause shallower UV transits.}
\label{fig:ratio_opacity}
\end{figure}

These opacities are displayed in Fig. \ref{fig:opac}.
One can see that small particles have much higher UV opacities while large particles
have almost flat (gray) opacities in UV and optical regions.
In Fig. \ref{fig:ratio_opacity} we plot the opacity ratios between opacities at 1300\AA\ and 5500\AA. One can indeed see that this ratio can be smaller than 1, and therefore the transit might be shallower in UV than in the optical region. In particular, this can happen for a broad range of iron grains ($-1.2 < \log r <0.0$ ) and mainly for silicate grains with $\log r \approx -0.7$.

\begin{figure}[ht]
\includegraphics[width=0.49\textwidth]{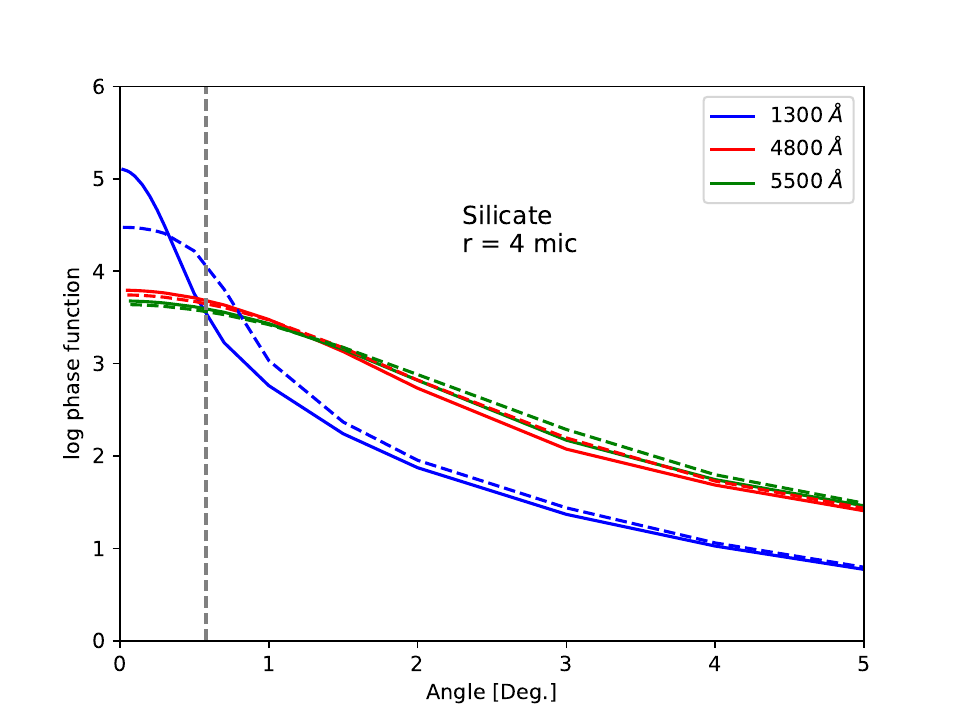}
\caption{Phase function of astronomical silicate dust particles with 
$r=4$ microns at 1300 \AA\ (blue line), 4800 \AA\ (red line), and 
5500 \AA\ (green line). The strongest forward scattering occurs at the 
shortest wavelength. Solid lines are normal phase functions while dashed lines are ``disk-averaged'' phase functions that take into account the finite dimension of the source of light. This is illustrated by a vertical line which is the angular radius of the star as seen by the grains.}
\label{fig:pf}
\end{figure}

Apart from opacities, phase functions might also cause even shallower UV transits.
The phase function describes the distribution of scattered photons as a function
of the scattering angle which is a deflection angle from the original direction of the photon.
A scattering angle of zero means that the photon continues in the original forward direction after the scattering event. Large grains at shorter wavelengths typically have a very 
pronounced and narrow forward scattering peak. This is illustrated in Fig. \ref{fig:pf} for
grains of astronomical silicate of 4 microns in size.
The importance of forward scattering in eclipsing systems during the primary eclipse, when
a cool dusty object is in front of the main source of light, was stressed in \cite{budaj2011} where it might cause a mid-eclipse brightening; it also
causes a pre-transit brightening observed in disintegrating exoplanets
\citep{rappaport12,brogi12,budaj13,lieshout16}.
It might easily also cause shallower UV transits, as photons in UV would continue in their original direction while those in the optical would be scattered to the sides and backward.
The ratio of the phase function (for the zero scattering angle) at 1300\AA\ and 5500\AA\
is shown in Fig. \ref{fig:ratio_opacity}. One can see that
this ratio is constant and is equal to one for small grains. This is because we are in the Rayleigh
scattering regime and phase functions at both wavelengths are the dipole phase functions
with roughly equal forward and backward scattering lobes.
For large values of $x=2\pi r/\lambda,$ the forward scattering peak increases and the phase function for non-polarized radiation at the zero angle is proportional to the extinction cross-section, which is proportional to $r^2$. As we increase the particle size, the phase function in UV enters this regime first
and the phase function ratio starts to increase. If we increase the particle size even more, the phase function in the optical enters this regime as well.
Consequently, both $r^2$ trends cancel out, and the phase function ratio becomes flat again and reaches values of about 15-20. This means that during the transit (zero scattering angle), more UV photons are scattered into the original direction and continue towards the observer in contrast to photons at longer wavelengths.
This demonstrates that the forward scattering regime can also be very important in this context of shallow UV transits. The importance of this effect 
also depends on the contribution of scattering to the total opacity.

Finally, we take into account a finite dimension of the source of light (star) as seen by the grains.
The angular diameter of the star is about $1.2\degr$, which is not large but is comparable
to or even larger than the width of the forward scattering peak for UV photons and larger grains.
We used the code DISKAVER and the method described in \cite{budaj2015} which takes this effect into account by calculating a ``disk-averaged'' phase function.
A comparison of these disk-averaged phase functions and normal phase functions is also shown in Fig. \ref{fig:pf}.

\subsubsection{Transit depth calculations}

Once we have the dust grain properties we calculate the transits using
our code Shellspec \citep{budaj04}.
The star is assumed to be a sphere with a radius $R_\star$ = 0.012$R_\odot$ and a mass of 0.6 $M_\odot$ \citep{izquierdo18}. Its spectrum, which forms the boundary condition for the radiative transfer in the CM, was taken from the previous chapter.
The star is subject to limb darkening. We applied a quadratic limb darkening law for DBA white dwarfs from \cite{claret20}.
The dust cloud is modeled using an object called RING predefined in the code.
RING is an arc (fraction of a ring) of dust located at a distance of 1.2 $R_{\odot}$ from the star.
For simplicity, we assume that its half-thickness in the vertical and radial directions is equal to 
the radius of the star. Dust density, $\rho_d$, is constant in the vertical and radial direction and decreases exponentially along the arc:
\begin{equation}
\rho_d(t)=\rho_{d}(t_0) e^{C},~~~~C=\frac{|t-t_0|}{\pi}D,
\label{}
\end{equation}
where $t,t_0$ are angles along the arc in radians and $D$ is the dust density exponent, respectively.
We assumed $D=-9$ which roughly fits the duration of transits. 
The light curves are calculated by rotating the line of sight (observer) assuming an inclination of $90\degr$. From light curves we calculate the transit depths and transit depth ratios (TDRs) for different wavelengths which we use for comparison with the observed values.
We calculate the transit depths for two species, iron and astronomical silicate,
and for a range of particle sizes 
$-2.0 \le \log r \le 1.0$ and for the wavelengths of 1300 \AA, 4800 \AA,
5500 \AA, and 45\,000 \AA. These correspond to the central wavelengths of photometric observations with HST/COS, Meyer, DCT, and Spitzer telescopes. 

The transit depth $d$ is calculated following \cite{xu2019} :
\begin{equation}
d=\frac{1}{p_2-p_1}\times \int^{p_2}_{p_1}f_0 - f(p)dp,
\label{D}
\end{equation}
where $p_1$ and $p_2$ represent the phase interval of interest,
$f_0$ is the out-of-transit flux and $f(p)$ is flux during the transit
as a function of phase $p$. This quantity depends on the choice of the interval but this dependence will drop out of the equation once the ratio at different wavelengths is calculated.

First we fit for the observed transit depths in the optical region.
Different dust densities are required for different dust particle sizes to fit the transit depth. 
In Table \ref{tab:dens} we list the approximate dust densities at the edge of the dust arc, $\rho_d(0)$, required for different particle sizes assuming the dust cloud geometry described above. Approximately two times more iron than silicate (by mass) is required to produce the same dip in the optical region,
except for very small grains. This means that the iron dust is slightly less optically active than silicate grains. The abundances of silicon, iron, and magnesium (main ingredients of silicates) are all very similar in the atmosphere. If the iron dust were to contribute significantly to the transit then most of the iron atoms would have to be in the iron grains (not in the silicates), or a significant fraction of magnesium or silicon would have to be in some other form of dust and not in the silicates, assuming that the dust cloud has the same chemical composition as the atmosphere.  
Once we have the dust cloud model that fits the optical transit depth, 
then we calculate the UV and IR transit depths using the same dust cloud model. Then we calculate the TDRs for different wavelengths. 

\begin{table}
\caption{Densities at the edge of the dust cloud required to fit the transit depth in the optical region for different particle sizes assuming a ring like geometry (see the text).} 
\label{tab:dens}
\centering
\begin{tabular}{lrr}
\hline\hline
$\log r \rm [mic]$  & $\rho_{\rm silicate} \rm [g\,cm^{-3}]$ & $\rho_{\rm iron} \rm [g\,cm^{-3}]$ \\
\hline
-2.0 & $2.2\times 10^{-13}$ & $2.5\times 10^{-14} $\\
-1.0 & $9.7\times 10^{-15}$ & $1.8\times 10^{-14}$\\
0.0  & $1.2\times 10^{-13}$ & $2.5\times 10^{-13}$\\
1.0  & $2.2\times 10^{-12}$ & $4.5\times 10^{-12}$\\
\hline
\end{tabular}
\end{table}

\begin{figure}[ht]
\includegraphics[width=0.49\textwidth]{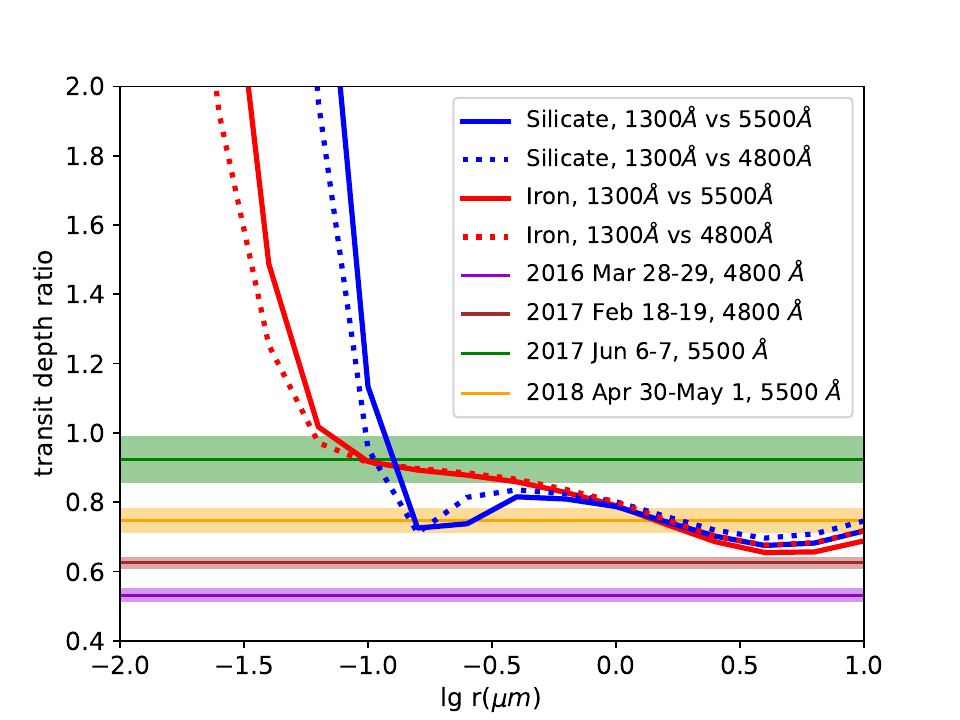}  
\includegraphics[width=0.49\textwidth]{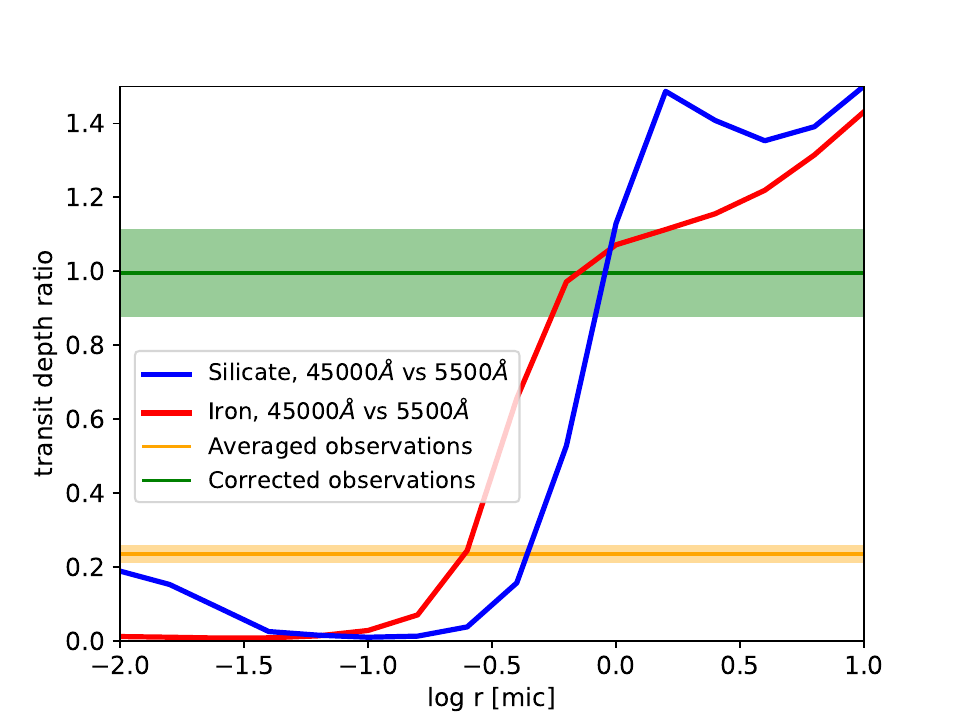}
\caption{Calculated transit depth ratio for silicate (blue) and iron (red) grains compared  with observations on different dates. Top: Ratio of 1300 \AA\ vs 5500 \AA\ (or 4800 \AA).
Bottom: Ratio of 45\,000\AA\ vs 5500\AA.
Horizontal colored lines correspond to the observed values. 
The widths of the semi-transparent horizontal colored lines correspond to 2 $\sigma$ errors. }
\label{fig:ratio}
\end{figure}

The results and their comparison with the observed TDRs of 1300\AA/5500\AA\ and 1300\AA/4800\AA\ from \cite{xu2019} are displayed in Fig. \ref{fig:ratio} (top panel). 
A TDR larger than one means that UV transits are deeper and we observe the well-known dust reddening. Values smaller than one mean that UV transits are shallower and we observe a bluing effect. One can see that particles smaller than about 0.1 microns cannot cause shallower UV transits because our synthetic TDRs are larger than 1 which is due to opacity ratios being greater than 1.
For silicates, we see a minimum in TDR at about $\log r=-0.8$, which is also caused
by the opacity ratio, and another broader and deeper minimum for $\log r \in (0.2,1.0)$, which is caused by the phase function ratio. 
For iron we observe TDR values slightly smaller than 1 for $\log r \in (-1,0)$
which are caused by the opacity ratio and even smaller values for $\log r \in (0.2,1.0)$
caused by the phase function ratio.
Therefore,  iron or silicate grains of 2-10 microns can naturally explain two observed shallow UV transits with TDR > 0.7.
However, we note that this model cannot explain two other UV transits with
the smallest TDR. This may be due to the limitation of our model, which considers only
optically thin scattering and single scattering events. It is possible that the dust cloud is more optically thick. This case would require a more sophisticated model including multiple scattering events and 3D radiative transfer. It is also possible that they are caused by the 
underlying gas ring absorption as suggested by \cite{xu2019}.
Optically thin dust clouds have to be more extended in the vertical direction compared to the
optically thick clouds to cause sufficiently deep eclipses. According to \cite{izquierdo18},
this would require unknown vertical motions or turbulence, the origin of which remains to be explained. 
In any case, an optically thick cloud will likely have optically thin edges. These edges will produce chromatic effects during the transit and the effects of opacity and mainly forward scattering we described above will also be important in this case.

Our synthetic TDRs for 45\,000\AA/5500\AA\ and their comparison with the observations of \cite{xu2019} are also shown in Fig. \ref{fig:ratio} (bottom panel).
We note that here we compare the ratio of a longer to a shorter wavelength, and so values smaller than one
represent the dust reddening and values larger than one the dust bluing. 
The yellow horizontal line is an average observed TDR from \cite{xu2019}. This value was corrected to account for an IR dust emission and is shown by the green line. 
These observations are best explained with moderate,
micron sized particles of both silicate and iron but larger particles might still be possible, especially if the cloud were optically thicker. 
A combination of TDRs from UV/optical and IR/optical regions indicates that particles of both iron and silicate of 
$1-10$ microns are the best candidates. This conclusion is based mainly
on simple dust grain properties. Such particles are relatively large 
and require larger dust densities. A smaller mass fraction of dust in the form
of smaller particles could produce a larger extinction (see Table \ref{tab:dens}).
Consequently, the question arises as to why small particles are missing. This might be because small particles
have higher equilibrium temperatures and evaporate much more quickly than larger particles as suggested by \cite{xu2018}.

\section{Variability of the circumstellar lines}
\label{variab}

\begin{figure}[ht]
\includegraphics[width=0.49\textwidth]{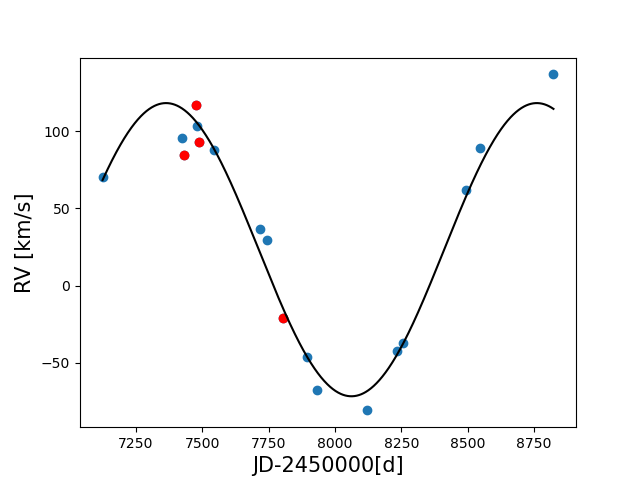}
\includegraphics[width=0.49\textwidth]{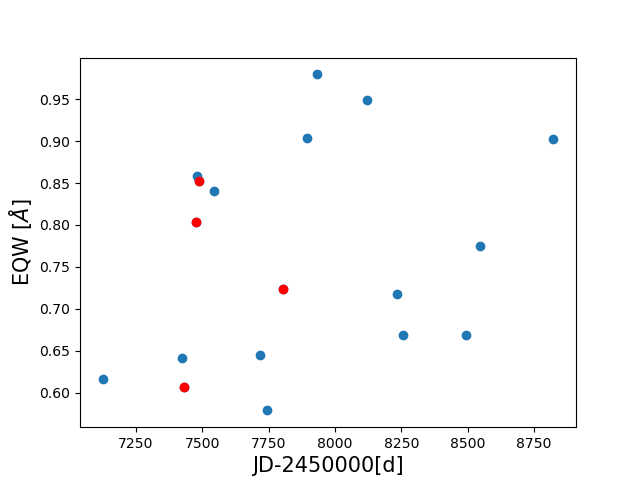}
\caption{Evolution of RVs (top) and EWs (bottom) of the \ion{Fe}{II} 5316 line. Observation from Keck (blue) and VLT (red) are shown.}
\label{rv_eqw}
\end{figure}

There are numerous circumstellar lines seen in the spectra.
Similarly to \cite{cauley18}, we chose the \ion{Fe}{II} 5316 line to study its variability 
because this line is seen in most of the spectra. In particular, we study the equivalent widths (EWs) and radial velocities (RVs) of this line.
The circumstellar component is a major contribution to the line and is blended by the stellar component. Because we use
data from different instruments (with different spectral resolutions), we decided not to remove the stellar component and instead measure the whole blend. This can slightly
underestimate the amplitude of the RV variations and overestimate
the EWs of circumstellar lines, but the outcome is model independent and
shows much less scatter. The results are shown in Fig. \ref{rv_eqw}
and Table \ref{rv_eqw_tab}.
One can see that the RVs have a clear periodic behavior and that observations cover the whole cycle.
We fitted a simple sine wave to the RVs:
\begin{equation}
RV= A + B \sin \left( \frac{2\pi(t-t_{0})}{P}\right),
\end{equation}
where $t$ is JD in days, RV is in km\,s$^{-1}$, $A=23.3\pm 4.1$ km\,s$^{-1}$, 
semi-amplitude $B=-94.9\pm 5.0$ km\,s$^{-1}$, $t_{0}=2457712\pm 14$, and period
$P=3.83 \pm 0.12$ yr. We use this ephemeris to calculate phases below.
The period is slightly shorter than previous estimates of 5.3 or 4.6 yr \citep{cauley18,fortin20}. Such a clear sine wave variability indicates that its nature is a robust physical phenomenon with little room for chaotic behavior, which supports the precessing elliptical disk model. 
Equivalent widths are also strongly variable due to changes in the circumstellar material.
These changes do not follow the periodic behavior of RVs and are apparently more sensitive
to other phenomena such as temperatures, densities, changes in inclination, vertical width, turbulence, warps in the disk, or transits of dust clouds.

\begin{table}[ht]
\caption{
Radial velocities in km\,s$^{-1}$ and EWs of the \ion{Fe}{II} 5316 line in \AA.}
\label{rv_eqw_tab}
\centering
\begin{tabular}{lllrr}
\hline\hline
Date  & Instrument & JD & RV & EQW\\
\hline
2015-04-11 &  HIRES     & 2457123 &   70.5 & 0.616 \\
2016-02-03 &  HIRES     & 2457421 &   95.4 & 0.641 \\
2016-02-14 &  XSHOOTER  & 2457432 &   84.4 & 0.607 \\
2016-03-29 &  XSHOOTER  & 2457476 &  116.9 & 0.803 \\
2016-04-01 &  HIRES     & 2457479 &  103.3 & 0.858 \\
2016-04-08 &  XSHOOTER  & 2457486 &   92.6 & 0.851 \\
2016-06-03 &  HIRES     & 2457542 &   88.0 & 0.840 \\
2016-11-26 &  HIRES     & 2457719 &   36.9 & 0.644 \\
2016-12-22 &  HIRES     & 2457745 &   29.7 & 0.578 \\
2017-02-20 &  XSHOOTER  & 2457804 &  -20.9 & 0.724 \\
2017-05-22 &  HIRES     & 2457895 &  -46.0 & 0.903 \\
2017-06-27 &  HIRES     & 2457931 &  -67.3 & 0.979 \\
2018-01-01 &  HIRES     & 2458120 &  -80.7 & 0.949 \\
2018-04-24 &  HIRES     & 2458232 &  -42.1 & 0.718 \\
2018-05-18 &  HIRES     & 2458256 &  -37.0 & 0.669 \\
2019-01-09 &  HIRES     & 2458493 &   62.1 & 0.669 \\
2019-03-02 &  HIRES     & 2458544 &   88.7 & 0.775 \\
2019-12-05 &  HIRES     & 2458823 &  136.7 & 0.902 \\
\hline
\end{tabular}
\end{table}

\section{Model of the gaseous disk}
\label{gasdisk}

\subsection{Previous models of the eccentric rings}
\label{oldringmodel}

\cite{cauley18} observed that the broad absorption features
shift from red to blue and proposed a model of 14 precessing
eccentric gas rings to explain this behavior. The eccentricities of the rings were 0.25-0.30
and periastra 19-28 $R_{\star}$. The precession rate of such disks due
to general relativity (GR) would be about 6 yr, which was close to the
observed period of red- to blueshifted cycles. Rings with smaller
periastra (<10 $R_{\star}$) and smaller eccentricities (<0.1) were excluded because the precession would be very fast.

This model was further improved by \cite{fortin20}. 
Their model also has 14 coplanar eccentric rings with similar
eccentricities. Their periastra range from 16 to 24
$R_{\star}$ and are gradually tilted in the true anomaly.
The whole structure precesses as a solid body with a 4.6 yr period. 
The precession is retrograde. The temperature of
the rings is about 6000 K but the model also includes a possible vertical
temperature stratification with midplane temperatures of up to 13\,000K. 
The midplane gas densities in the rings are about $6\times 10^{-6}$ g\,cm$^{-3}$. 
Given the radial thickness of these rings (0.5 $R_{\star}$), such a density would correspond to column densities of the material transiting the star in excess of $10^1$ kg\,cm$^{-2}$.
Such gas rings would be quite opaque and might significantly alter the SED.
The total mass of rings would be about $10^{23}$ g or $10^{-5}$
$M_{\oplus}$ (we note that this is significantly higher than
$2\times 10^{16}$ g claimed by the authors).
The model involves real opacity
calculations and absorption produced by the disk but neglects the line emission. It can indeed very well
reproduce the shape of most spectral lines from UV to the optical region and their temporal evolution. However, there are still certain limitations and disadvantages 
to the model, most of which were pointed out by the authors themselves. 
We attempt to resolve some of them with our model, which is described in the following sections.

\subsection{A modified or alternative model of the eccentric disk}

We used the Shellspec code to test new models of gas rings.
The code has many predefined 3D structures and objects and the user can compose
his or her own model from these objects and calculate their spectra and light curves.
Unfortunately, there was no object which would suit our needs, and so we modified the Shellspec code and implemented a new 
object called ELLIPSE, which is a noncircular (elliptical) disk-like structure around 
an object with a mass of $M_{\star}$ located at the center of the coordinate system, which is a focus of the ellipse.
The disk is characterized by a single value of eccentricity, $e$, which is
the same for all stream lines.
In the orbital plane, it is limited by an inner and outer ellipse 
with radius at the periastron $p_{\rm in}, p_{\rm out}$, respectively.
In the vertical direction it is limited by a few pressure scale heights given by the $ae$ parameter. The object may have a net space velocity.
A physical justification for the existence of such eccentric gas rings around white dwarfs was presented recently by \cite{trevascus21}. Using smoothed particle hydrodynamics, the authors demonstrated that a sublimating or partially disrupting planet on an eccentric orbit around a white dwarf would form and maintain a gas disk with 
an eccentricity similar to or lower than that of the orbiting body.

ELLIPSE has an optional inclination but in the following we describe its properties
in the cylindrical coordinates $(r,\theta,z)$ aligned with the disk.
We incorporated two different radial temperature behaviors.
The first is an empirical radial temperature law at the disk plane:  
\begin {equation}
T(r)=T_{0}(r/p_{\rm in})^{etmp} ,
\label{tdisk2} 
\end {equation}
where $T_{0}$ is the temperature at the inner disk radius $p_{\rm in}$.
Such behavior is typical for irradiated disks with a typical value of $etmp=-0.5$.
The second behavior is the standard radial stratification:
\begin {equation}
T(r)=T_{0}\left(\frac{R_{\star}}{r}\right)^{\frac{3}{4}}
\left(1-\sqrt{\frac{R_{\star}}{r}}\right)^{\frac{1}{4}}.
\label{tdisk3}
\end {equation}
Here, $T_{0}$ is the characteristic disk temperature \citep{pringle81}
and $R_{\star}$ is the radius of the central object. 
Such temperature behavior is expected for an optically thick steady-state accretion disk
in a strong gravitational field radiating at the expense of the potential energy of the accreted material. Maximum disk temperature is achieved slightly away from 
the stellar surface and is about half of the characteristic disk temperature.
$T_{0}^4$ is proportional to the mass accretion rate.

Vertical density profile and the extent of the disk is controlled
by its scale height, H, which is a function of radius,
velocity, $v$, and the speed of sound, $c_{\rm s}$:
\begin {equation}
H(r)=h_{\rm e} \frac{c_{\rm s}(r)}{v(r)}r.
\label{scale1}
\end {equation}
Such behavior is expected for steady state circular accretion disks.
As our disk is elliptical, we introduced
an extra free parameter, $h_{\rm e}$, that might be adjusted to fit the observations,
expand the pressure scale height, and mimic the effect of turbulence  for example.
For this sole purpose, we assume that the velocity is a circular
Keplerian velocity:
\begin {equation}
v(r)=\sqrt{G\frac{M_{\star}}{r}},
\label{scale2}
\end {equation}
where $G$ is the gravitational constant.
The speed of sound is
\begin {equation}
c_{\rm s}(r)=\sqrt{\gamma k T(r)/\mu},
\end {equation}
where $\gamma=5/3, 7/5, 1$ is an adiabatic index for a monoatomic, 
diatomic, or a more complicated molecule with many degrees of freedom, 
respectively,
$k$ is the Boltzmann constant, and $\mu$ is the mean molecular weight
calculated from the chemical composition:
\begin{equation}
\mu=\frac{\Sigma a_{i} m_{i}}{\Sigma a_{i}},
\end{equation}
where $a_{i}, m_{i}$ are the element abundances and their atomic masses, respectively.

For any given location (x,y,z), we calculate $r=\sqrt{x^2+y^2}$
and a true anomaly, $\theta$, from the interval $<-\pi,\pi>$:
\begin {equation}
\begin{array}{ll}
\theta= \arccos(x/r) ~~ & {\rm for}~ y>0,  \\
\theta=-\arccos(x/r) ~~ & {\rm for}~ y<0.
\end{array}
\end {equation}
The eccentric anomaly, E, from the same interval $<-\pi,\pi>$ is
\begin {equation}
E=2 \arctan \left( \sqrt{\frac{1-e}{1+e}} \tan\frac{\theta}{2} \right).
\end {equation}
Using the eccentric anomaly, the components of the Keplerian velocity 
vector $\vec{v}$ are
\begin {equation}
\begin{array}{l}
v_x=- \sin E ~\sqrt{GM_{\star}a}/r              \\
v_y=\sqrt{1-e^2}\cos E ~\sqrt{GM_{\star}a}/r    \\
v_z=0~ .
\end {array}
\end {equation}
The semi-major axis $a$ is
\begin {equation}
a=r\frac{1+e\cos\theta}{1-e^2}.
\end {equation}
It is possible to add an extra radial velocity component to mimic 
outflows or inflows of the material.

We assume that the surface gas density at the periastron , $\Sigma_p$, varies along the line of apsides (i.e., with the periastron distance from the center) as a power law:
\begin {equation}
\Sigma_{p}=\Sigma_{p_{\rm in}}(p/p_{\rm in})^{eden} ,
\end {equation}
where $p_{\rm in},\Sigma_{p_{\rm in}}$, and $eden$ are the inner radius (periastron) 
of the disk, a surface density at that point, and a density exponent, respectively.
Such behavior is expected for accretion disks. Unfortunately, the density exponent
is not very well constrained. For example, studies of Be star disks indicate 
a very broad range in midplane density exponent (-1.5,-4) \citep{silaj10}, while a recent detailed analysis of interferometric, spectroscopic, and photometric observations of $\beta$ Lyr revealed a value of $eden=-0.6$ for this system \citep{broz21}.
Surface gas density $\Sigma_{r}(\theta)$ at some true anomaly $\theta$ and
distance $r$ corresponding to some periastron $p$ can be obtained from 
the continuity equation:
\begin {equation}
\Sigma_{r} |\vec{r}\times\vec{v}(r)| dr/r=\Sigma_{p}v(p)dp .
\label{sigmarp}
\end {equation}
The radius distance for an ellipse with periastron oriented along the x-axis 
is
\begin {equation}
r= \frac{p(1+e)}{1+e\cos\theta}.
\end {equation}
Assuming a constant value of eccentricity, the change in radius corresponding to a change in periastron is
\begin {equation}
dr= \frac{(1+e)}{1+e\cos\theta} dp.
\label{drdp}
\end {equation}
The vector cross product is
\begin {equation}
|\vec{r}\times\vec{v}(r)| = p v(p).
\label{rxv}
\end {equation}
%
%
Combining Eqs. (\ref{sigmarp}), (\ref{drdp}), and (\ref{rxv}), 
we obtain:
\begin {equation}
\Sigma_{r}(\theta) = \Sigma_{p} .
\end {equation}
%
%
Surface densities along the streamline are constant because of the second Kepler's law (Elliot Lynch priv. comm., \citealt{ogilvie01})

We assume the Gaussian vertical density distribution
typical for steady-state gas-pressure-do\-mi\-na\-ted accretion disks:
\begin {equation}
\rho(r,z)=\rho(r,0)\exp\left(-\frac{z^2}{2H^2}\right) .
\end {equation}
Its midplane gas density, $\rho(r,0)$, is determined (as a function of 
the distance) from the surface density and the vertical scale height, 
$H$, using the relation
\begin {equation}
\Sigma_r=\sqrt{2\pi}H(r)\rho(r,0).
\label{sigmah}
\end {equation} 
%
The electron number density is calculated from the gas density, temperature, and chemical composition assuming LTE.
We allow for an extra turbulence of the material specified by the parameter $vtrb$.
The motivation for this parameter comes from the definition of microturbulence
in stellar atmospheres. It is a movement on scales shorter than the mean free path of a photon. We add it to the thermal line broadening when calculating the Voigt profile.
It may mimic departures from the Keplerian velocities or desaturation of spectral lines due to 
the velocity gradients in combination with a finite step in our 3D space grid.
The velocity field can easily change by more than 10 km\,s$^{-1}$ from one grid point to another and such turbulence also eliminates the associated numerical noise.
In addition, we modified the Shellspec code to take into account the gravitational redshift of the stellar surface. It was assumed that the circumstellar material is at zero redshift, which is a reasonable assumption \citep{fortin20}.
Finally, we also modified the code to use the thermal (true) opacity from a pre-calculated table.
The table in the vicinity of the \ion{Fe}{II} 5316 line was calculated using the code SYNSPEC \citep{hubeny21}. Scattering opacities were still calculated in the Shellspec code.

\subsection{Results, disk parameters, and discussion}

Our model consists of a star and gaseous circumstellar material. 
The star was modeled as a sphere with a radius and mass of $0.012 R_{\odot}$ and $0.6 M_{\odot}$, respectively. It is subject to limb darkening and we adopted the quadratic limb darkening law. The corresponding coefficients of the limb darkening
were taken from \cite{claret20}. The intrinsic spectrum of the star was taken from Sect. \ref{atm}. The circumstellar material has the same chemical composition as the star but we excluded helium because its abundance in rocky planets or small Solar System bodies is negligible. We then experimented with the above-mentioned disk model and its parameters so that it would fit the \ion{Fe}{II} 5316 line profile and its variability. Unfortunately, such a disk with fixed eccentricity cannot reproduce this line very well. However, adding a second disk with a different eccentricity improves the situation considerably.
We also experimented with different radial temperature behaviors. We find that the effective temperature behavior given by Eq. (\ref{tdisk3}), which applies to opaque steady accretion disks radiating at the expense of their potential energy, is too steep  and does not reproduce the observations very well.
Most probably, this is because our disk is optically thin and eccentric, and is therefore far from the assumptions of Eq. (\ref{tdisk3}). Gas temperatures in such a disk are governed by heating and cooling balance and may be different from the effective temperature \citep{hubeny90}. It has also been noticed that viscous heating within the concept of a standard accretion disk cannot reproduce observed emission lines from such disks either \citep{hartmann11}.  That is why we adopted the temperature structure according to Eq. (\ref{tdisk2}).

\begin{figure}[ht]
\includegraphics[width=0.49\textwidth]{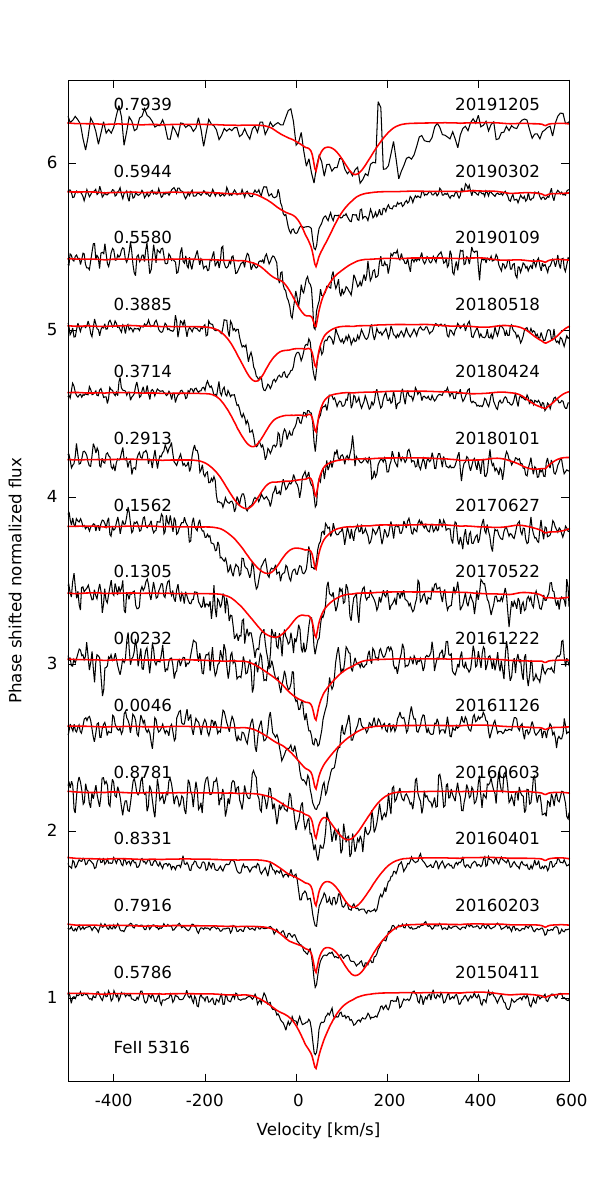}
\caption{Comparison of the observed \ion{Fe}{II} 5316 line profile (black line) and the theoretical spectrum 
(red line). Labels on the right denote year, month, and day of the observations
while labels on the left denote the phase. Phase zero roughly corresponds to the transit of the apoastra.}
\label{fig:feii}
\end{figure}

\begin{figure}[ht]
\includegraphics[width=0.49\textwidth]{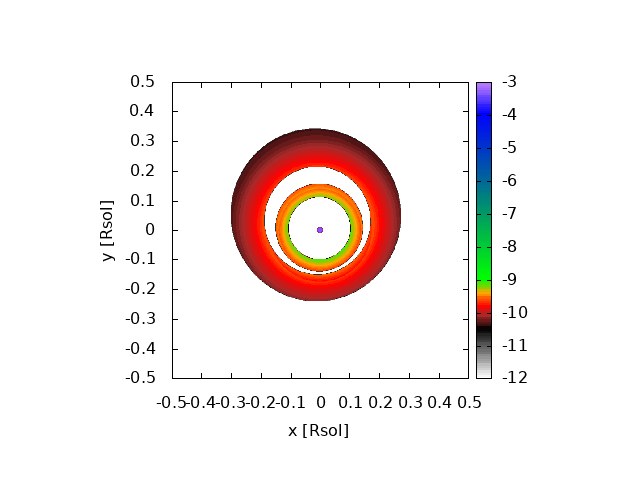}
\caption{View of the disks from above. 
Zero phase corresponds to looking from the positive y-axis direction.
The colors represent a logarithm (base 10) of intensity at 5300 \AA\
in $\rm erg\,cm^{-2}\,s^{-1}\,Hz^{-1}\,sterad^{-1}$.} 
\label{fig:2D}
\end{figure}

We converged to the following model with two disks:
an inner disk, which is slightly hotter and almost circular, and an outer disk, which is cooler and more eccentric. We assume that the orbital planes of both disks are aligned and we see them edge on 
($i=90 \degr$). The whole structure precesses as a solid body.
The parameters of the disks are given in Table \ref{tab:disk}, their geometry is illustrated
in Fig. \ref{fig:2D}, and the comparison of the observed and theoretical spectrum of this \ion{Fe}{II} line is shown in Fig. \ref{fig:feii}.

We set the exponent of the temperature behavior of both disks ($etmp$) to be identical.
After some experiments with this parameter, its value turned out to be relatively low. Temperatures in the inner disk span the region of 5700-5200 K and the outer disk
has temperatures in the range of 5500-4700 K. This is approximately in agreement with the findings of previous authors \citep{fortin20} who used a zero radial gradient.
We fixed the density exponents ($eden$ parameters) of both disks to the same value and fitted their inner midplane densities.
It turned out that these densities ($\rho (p_{\rm in},0)$) are very similar. This is not coincidental; it indicates that although we treat both disks independently, they form a single smooth coherent structure which might be understood as a single disk with smoothly varying eccentricity. Lines of apside of both disks coincide. However, these lines are slightly
tilted with respect to the phase zero direction which is defined by the RV curve.
This tilt is small, about 15 degrees, and so the phase zero roughly corresponds to the transit of the apoastra.
The fit to the observations is reasonably good given the relatively low number of free parameters
and observed variability of EWs.

The most significant problem is enhanced absorption in the model during transits of disk periastra (phases near 0.5). This may be caused by the vertical scale height 
and Equations \ref{scale1} and \ref{scale2}. These are expected for steady circular disks and may not describe this scale height properly in the case of eccentric disks. The scale height increases with radius and this dependence might be too steep. We boosted the vertical scale height of our disks using the $h_{\rm e}$ parameter by factors of ten and four for the inner and outer disks, respectively. Such a boost might be justified by a turbulence, perturbations, or by the fact that our disks are far from  being ideal steady, circular accretion disks. We also used a turbulence of 20 km\,s$^{-1}$.
This value might appear large and supersonic, but, again, the real disks may experience various inflows, outflows, precession, and interactions between the two disks (mainly near periastron). Given that the Keplerian velocities are of the order of 1000 km\,s$^{-1}$, these are relatively small departures from purely Keplerian orbits. Problems near periastron might also be caused by a small inclination or misalignment between the two disks, which might preferably boost absorption at other phases.

 Assuming the averaged parameters of the outer disk ($a=0.24 R_{\odot}, e=0.18$) and the mass of the star (0.6 $M_{\odot}$), GR would predict a precession rate of about 3 yr, which is roughly in agreement with the observed period of RV variability. Precession of the inner disk
due to GR would be much shorter (about 0.6 yr). However, this is not a significant issue in our model because the inner disk has a very low eccentricity and contributes much less to the RV variability than the outer disk. This may also be the reason for the poorer fit at some moments and might trigger the variability in EWs.
We note that our model is precessing in the prograde direction (transit of the apoastron is followed
by a blueshifted absorption) which is also in agreement with GR theory.
The total mass of both disks is $5\,10^{17}$ g or $3\,10^{-16} M_{\odot}$ or $8\,10^{-11} M_{\oplus}$. This value is an order of magnitude higher than our estimate of the minimum amount of metals in the convective zone ($2\,10^{-11} M_{\oplus}$).

\begin{figure}[ht]
\includegraphics[width=0.49\textwidth]{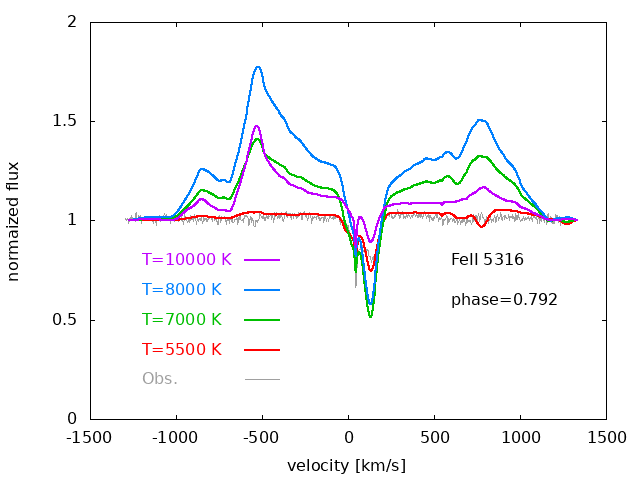}
\caption{Line profiles of \ion{Fe}{II} 5316 for different inner disk temperatures.
A clear double peak emission emerges in hotter disk models.}
\label{fig:emission}
\end{figure}

An important limitation on the temperature of the disk comes from the line emission.
This is why we cannot significantly raise the disk temperatures.
To illustrate this effect, in Fig. \ref{fig:emission} we plot the line profiles
for disks with different $T_{0}$ temperatures. For simplicity we assume that the densities have not changed and that $T_{0}$ has the same value for both disks. The line opacity and emissivity increase with disk temperature, reach a maximum near 8000K, and then start to decline.
One can see a clear double peak emission typical of accretion disks. 
The model was calculated near the quadrature (phase of 0.792). The blue emission is created by the material near the apoastra while the red emission peak is due to the material near the periastra and is shifted to higher velocities.

\begin{table}[ht]
\caption{Parameters of the circumstellar disks.}
\label{tab:disk}
\centering
\begin{tabular}{lll}
\hline\hline
Param.                    &  Disk 1    &  Disk 2 \\
\hline
$p_{\rm in} [R_{\odot}]$  &  0.10      &  0.15  \\
$p_{\rm out} [R_{\odot}]$ &  0.14      &  0.24  \\
e []                      &  0.06      &  0.18  \\
ae [H]                    &  3.        &  3.    \\
$h_{\rm e}$ []            &  10.       &  4.    \\
i [deg]                   &  90.       &  90.   \\
$\alpha$ [deg]            &  -15.      &  -15.  \\
$T_{0}$ [K]               &  5700      & 5500   \\
etmp []                   &  -0.2      & -0.2   \\
$\rho (p_{\rm in},0) \rm [g/cm^3]$ & $3.3\,10^{-12}$ &  $2.5\,10^{-12}$ \\
eden []                  & -1.        & -1.    \\
vtrb [km/s]              &  20.       &  20.   \\
$R_{\star} [R_{\odot}]$  & 0.012      & 0.012  \\
$M_{\star} [M_{\odot}]$  & 0.63       & 0.63   \\
\hline
\end{tabular}
\end{table}

This model has the following advantages when compared to the previous models described
in Sect.\ref{oldringmodel}:

- a smaller number of objects (2 disks versus 14 rings);

- radial temperature gradient was considered;

- our model preserves the continuity equation of the flow; 

- line emission from the disks is taken into account and it poses a significant limitation for the temperature of the disks;

- a slightly better fit during the transit of the apoastron and of the most blueshifted profiles but poorer fit just after and during the transit of the periastron;

- the precession of our model is prograde which is in agreement with the precession following GR theory;

- the total mass of the whole structure is significantly different ($8\,10^{-11} M_{\oplus}$ of our model in comparison to $10^{-5}$ $M_{\oplus}$). 

The main disadvantage of our model is that it is based on only one spectral region while
the previous model includes the whole spectrum from UV to the optical region.
At present, it is rather cumbersome to extend these calculations to longer spectral intervals
because they involve the whole volume of the accretion disks and are relatively slow in comparison to previous models.
We also did not consider the vertical temperature dependence because it would be hard to justify in the present model.
It should be stressed that our model is certainly not the only possible model or 
the best. It should be understood only as an example of another possibility
or an alternative to previous models, and can certainly be improved upon  further.

\section{Conclusions}

In this paper, we offer alternative models of the atmosphere, gas rings, and dust clouds of WD1145.  Our main results are:
\begin{itemize}
\item
We calculated LTE, deep LTE, and NLTE atmosphere models of the star and synthetic spectra
using the codes TLUSTY and SYNSPEC. This is probably the first time that these codes were used to study such a heavily polluted, helium-dominated white dwarf (DBZA class). The models include line blanketing and \ion{He}{I} level dissolution using formalism developed by \cite{hubeny94}.
\item
We estimated the mass of the upper radiative layers to be about $5\,10^{18}$ g or $9\,10^{-10} M_{\oplus}$ and put a lower limit on the mass of the convection zone of $2\,10^{21}$ g
or  $3\,10^{-7} M_{\oplus}$. The mass fraction of metals is $Z=6\,10^{-5}$ and so the convection zone
contains at least $2\,10^{-11} M_{\oplus}$ of metals.
\item
We determined the chemical composition using a restricted NLTE approach and NLTE spectrum synthesis. Our metal abundances are slightly lower  
then previous findings but are in good agreement with them when expressed relative to silicon abundance. The chemical composition of the atmosphere is similar to that of CI chondrites except for carbon, nitrogen, and sulfur which are significantly underabundant. These key elements  
resemble rather a rocky planet (Earth). This provides strong support for the idea that the star was recently hit by an object of Earth-like composition, as suggested by \cite{fortin20}.
\item 
Some shallow UV transits observed in WD1145 can be naturally caused by the optical properties of dust grains, by their opacities, and mainly by strong forward scattering.
Forward scattering is much stronger and has a narrower peak in UV than in the optical region. This leaves more UV photons traveling in the original direction during the transit. Grains of silicate and iron of 1-10 microns in size are the best candidates. 
This can explain the two shallow UV transits of \cite{xu2019}.
Two other, even shallower transits might require more optically thick dust and
more sophisticated 3D radiative transfer modeling or some other explanation.
For example, the underlying gas ring absorption as suggested by \cite{xu2019}.
However, the effects described above may also play a crucial role in those cases.
\item
Radial velocity of the circumstellar lines shows a clear sine wave behavior with a period of 3.83 yr. Equivalent widths of circumstellar lines are also variable but they do not follow the variability in RV and their variability is caused by other phenomena.
\item
We created an alternative model of a gas cloud surrounding the star which can explain the behavior of the circumstellar absorption. It is composed of two disks:
an inner, hotter, and almost circular disk, plus an outer, cooler, and eccentric disk.
The advantages and disadvantages of this model are summarized at the end of the previous section.
\end{itemize}

As a final note, we stress that in the present study we treated dust clouds and gas disks separately. 
In reality, their mutual interplay and location will be important \citep{izquierdo18,xu2019}
and deserves further investigation, but taking these into account is beyond the scope of the present paper.

\section{Acknowledgement}
We would like to thank two anonymous referees for a number of comments which led to significant improvements to the manuscript.
We also appreciate consultations on various topics with Elliot Lynch and our colleagues Miroslav Kocifaj, Oleksandra Ivanova, Lubo\v{s} Neslu\v{s}an, and J\'ulius Koza.

This research has made use of the Keck Observatory Archive (KOA), which is operated by the W. M. Keck Observatory and the NASA Exoplanet Science Institute (NExScI), under contract with the National Aeronautics and Space Administration. The following PIs of datasets are kindly acknowledged: Jura, Redfield, Debes, Zuckerman, Hillenbrand, Steele, and Xu. 
Based on data products from observations made with ESO Telescopes at the La Silla Paranal Observatory under programme IDs 296.C-5014, 598.C-0695.
Based on observations made with the NASA/ESA Hubble Space Telescope, obtained from the data archive at the Space Telescope Science Institute. STScI is operated by the Association of Universities for Research in Astronomy, Inc. under NASA contract NAS 5-26555 (PI Xu).
This work has made use of data from the European Space Agency (ESA) mission
{\it Gaia} (\url{https://www.cosmos.esa.int/gaia}), processed by the {\it Gaia}
Data Processing and Analysis Consortium (DPAC,
\url{https://www.cosmos.esa.int/web/gaia/dpac/consortium}). Funding for the DPAC
has been provided by national institutions, in particular the institutions
participating in the {\it Gaia} Multilateral Agreement.
The authors were supported by the VEGA 2/0031/22 and APVV 20-0148 grants.

\bibliographystyle{aa} 
\bibliography{my,budaj2}

\end{document}